\documentstyle[12pt,fleqn]{article}
\def\Omeg{{\mit\Omega}}
\def\Ph{{\mit\Phi}}
\def\Ps{{\mit\Psi}}

\begin{document}

\renewcommand{\thefootnote}{\fnsymbol{footnote}}
\newcommand{\beq}{\begin{equation}}
\newcommand{\eeq}{\end{equation}}
\newcommand{\0}{\mbox{$|0\rangle$}}
\newcommand{\1}{\mbox{$|1\rangle$}}
\newcommand{\2}{\otimes}
\newcommand{\zz}{\mbox{$|z\rangle$}}
\newcommand{\ao}{\mbox{$|a=0\rangle$}}
\newcommand{\ket}{\rangle}
\newcommand{\bra}{\langle}

\vspace*{16mm}
\begin{center}{\Large Error Correction and Symmetrization\\[2mm]
in Quantum Computers}\\[12mm] 
{\large Asher Peres}\footnote{Permanent address: Technion---Israel
Institute of Technology, 32\,000 Haifa, Israel}\\[7mm] {\sl
Institute for Theoretical Physics, University of California, Santa
Barbara, CA 93106}\end{center}\vfill

\noindent{\bf Abstract}\medskip

Errors in quantum computers are of two kinds: sudden perturbations to
isolated qubits, and slow random drifts of all the qubits. The latter
may be reduced, but not eliminated, by means of symmetrization, namely
by using many replicas of the computer, and forcing their joint
quantum state to be completely symmetric. On the other hand, isolated
errors can be corrected by quantum codewords that represent a logical
qubit in a redundant way, by several physical qubits. If
one of the physical qubits is perturbed, for example if it gets
entangled with an unknown environment, there still is enough
information encoded in the other physical qubits to restore the logical
qubit, and disentangle it from the environment.  The recovery procedure
may consist of unitary operations, without the need of actually
identifying the error.\vfill\newpage

\renewcommand{\thefootnote}{\arabic{footnote}}
\setcounter{footnote}{0}

\noindent{\bf 1. Introduction}\bigskip

A computer is a physical system, subject to the ordinary laws of nature.
No error ever occurs in the application of these laws. What we call an
error is a mismatch between what the computer does and what we wanted it
to do. This may be caused by incorrect programming (software errors,
that I shall not consider here), or by imperfect hardware. The computer
engineer's problem is to design the hardware in such a way that common
flaws, which are unavoidable, will almost never cause errors in the
final output (namely, in the relevant parts of the final state of the
computer).

In {\it classical\/} computers, logical bits, having values 0 or 1, are
implemented in a highly redundant way by bistable elements, such as
magnetic domains. The bistability is enforced by coupling the physical
bits to a dissipative environment. Errors may then occur, because
of thermal fluctuations and other hardware imperfections. To take care
of these errors, various correction methods have been developed [1],
involving the use of redundant bits (that are implemented by additional
bistable elements).

In {\it quantum\/} computers, the situation is more complicated: in
spite of their name, the logical ``qubits'' (quantum binary digits) are
not restricted to the discrete values 0 and 1. Their value can be
represented by any point on the surface of a Poincar\'e sphere.
Moreover, any set of qubits can be in an {\it entangled\/} state: none
of the individual qubits has a pure quantum state, it is only the state
of all the qubits together that is pure [2].  The continuous nature of
qubit states implies that there can be no intrinsic stabilizing
mechanism, and error control becomes critical.

Here, a distinction must be made between quantum computers of the
Benioff type~[3,~4], where quantum hardware is used for
implementing classical logic, and computers that are fundamentally
quantal [5], and can do more than just mimicking classical
computation. In the former case, there are instants of time at which all
the qubits ought to represent definite values, 0 or 1. They are not then
in a quantum superposition, and error correction can be done as for a
classical computer [6]. On the other hand, in a computer of the
Deutsch type [5], the quantum state of the computer typically
is an entangled state of all the qubits, and classical methods of error
correction are not applicable. What can be done then depends on the
nature of the anticipated errors.

In general, we may write the Hamiltonian of the computer as $H=H_0+H_1$,
where $H_0$ is the Hamiltonian of an ideal error free computer, and
$H_1$ represents the influence of the environment. The latter is unknown
to the computer designer, except statistically. That Hamiltonian acts
on a Hilbert space which is the tensor product of those representing the
computer and the environment. The designer's problem is to distill, from
the computer's variables, a subset giving with probability close to 1
the correct result of the computation, irrespective of the unknown form
of $H_1$ and of the initial state state of the environment. Two
different types of errors ought to be considered: accidental large
disturbances to isolated qubits (e.g., a residual gas molecule may hit
one of them), and small, random, uncorrelated drifts of all the qubits.

The first type of error can be corrected by using codewords, as first
shown by Shor~[7]. A codeword is a representation of a logical
qubit by means of several physical qubits. There were 9 qubits in Shor's
codewords. It is now known that the minimal number is~5. In particular,
Bennett {\sl et al.\/} have constructed 5-qubit codewords that have the
remarkable property of being invariant under a cyclic permutation of the
qubits [8]. In all these quantum codewords, the physical qubits
are in a highly entangled state, chosen in such a way that, if any one
of the qubits gets entangled with an unknown environment, there still is
enough information stored in the other qubits to restore the codeword
and to unitarily disentangle it from the environment, irrespective of
the unknown state of the latter.

The second type of error, continuous random drifts of all the qubits,
cannot be eliminated by using codewords, but can be reduced by
symmetrizing the joint quantum state of several identical computers
[9,~10]. This symmetrization method and some of its variants are
discussed in Sect.~2. The encoding of logical qubits into codewords
consisting of several physical qubits is explained in Sect.~3. How to
actually restore the initial state of a corrupted codeword is shown in
Sect.~4. Finally, Sect.~5 is devoted to a fundamental issue: error
correction without detection of the error syndromes.\\[7mm]

\noindent{\bf 2. Symmetrization}\bigskip

In its original version~[9,~10], the symmetrization method involved the
use of $R$ identical replicas of the entire computer. At preset times,
the joint quantum state of the $R$ computers is projected onto the
symmetric subspace of their common Hilbert space (for example, by
measuring whether or not the state is symmetric, and aborting the
computation if the answer is negative). As shown below, if small
errors randomly affect all the qubits, this projection reduces the
average error by a factor $R$.  On the other hand, symmetrization gives
poor results if a single qubit goes completely astray: we then have a
non-symmetric state that is almost orthogonal to the symmetric
subspace, and the computation is almost always aborted. Indeed, if one
of the computers has a state orthogonal to that of all the others, the
probability is only $1/R$ that the joint state will be projected onto
the symmetric subspace, and in that case, the error is not eliminated,
but rather uniformly spread over all the $R$ computers! Obviously,
large isolated errors cannot be handled in this way. Their correction
requires the use of specially designed codewords.

There is however a more efficient protocol for error correction by
symmetrization. The $R$ computers can be arranged in pairs, and each one
of the $R/2$ pairs is symmetrized separately. The process can then be
repeated with different pairing arrangements, if we wish to further
improve the symmetry. With such a pairwise symmetrization, if a computer
accidentally gets into a state orthogonal to that of all the other ones,
there is a 50\% chance that the pair containing the bad computer will be
eliminated, and a 50\% chance that the error will be equally shared by
the two computers. Repeating this process many times, so that each
computer has many partners, ultimately leads to the elimination of a
bad computer, together with one good one. There still are $R-2$ good
computers available for continuing the work.

A more complicated (and probably more realistic) model would be to assume
that any computer may occasionally fail when one of the logical steps is
executed. This event must be rare enough so that the total probability
of failure of any given computer during the entire computation is less
than $1\over2$. Pairwise symmetrizations are performed between any two
logical steps (the pairs are chosen in such a way that each computer is
compared with many other ones during the complete computation). Most
errors are then eliminated, and the surviving computers contain, on
the average, less than one defective result.  In this theoretical model,
an ``error'' means a state that is orthogonal to the correct one. This
has to be generalized to the case of less radical errors. It is
plausible that repeated pairwise symmetrizations are in general
preferable to a single overall symmetrization, but a formal proof is
still needed.

Clearly, the poor efficiency of the symmetrization method, in the case
of large isolated errors, is due to possible failures of the symmetry
tests (also known as ``quantum measurements''). When a test fails, we
must discard a pair of computers, if not the entire process. However,
there is no need of measuring anything in order to force a quantum
state to stay in a symmetric subspace. A measurement is not a
supernatural event. It is an ordinary dynamical process, and any error
correction that may result from it should also be obtainable as a
consequence of a unitary evolution, governed by ordinary dynamical
laws. Indeed, a much simpler method for enforcing symmetry of the
quantum state is to impose on the $R$ computers an extra static
potential that vanishes in the symmetric subspace, and has a very large
value in all the orthogonal (asymmetric) states. Effectively, in the
$R$ computers, any $R$ homologous physical qubits behave as if they
were $R$ bosons. Likewise, if the qubits of a codeword have an internal
symmetry, such as the cyclic symmetry of the codewords in ref.~[8], we
may protect their cyclic subspace by erecting around it a high
potential barrier.

The result of such a symmetrizing potential is analogous to a
continuous Zeno effect (ref.~[2], pp.~392--400). To test its
effectiveness, consider the simple example of two computer memories,
each one consisting of a single qubit, initially in the state
$\alpha\choose\beta$, which is unknown. We want these computer memories
to be stable:  there should be no evolution of the two qubits. The
problem is to protect them against random fluctuations of the
environment.  Let us use for this discussion the terminology and
notations appropriate to spin-$1\over2$ particles. A symmetric state of
the pair belongs to the triplet $(J=1)$ representation, while the
singlet $(J=0)$ is antisymmetric.

Consider the Hamiltonian

\beq H_0=(1-{\bf J}^2/2)\,\Omeg, \label{Om} \eeq
where $\Omeg$ is a large positive constant. Since ${\bf J}^2=J(J+1)$,
this potential vanishes in the triplet state, and is equal to $\Omeg$
for a singlet. As a simple model of perturbation, let a phase error be
generated by

\beq H_1=\mu\,\sigma_{Az}+\nu\,\sigma_{Bz}, \eeq
where $\mu$ and $\nu$ are constant coefficients much smaller than
$\Omeg$, and the subscripts $A$ and $B$ refer to the two qubits. This
can also be written as

\beq H_1=\epsilon\,(\sigma_{Az}+\sigma_{Bz})+
 \eta\,(\sigma_{Az}-\sigma_{Bz}), \eeq
where $\epsilon=(\mu+\nu)/2$ and $\eta=(\mu-\nu)/2$. The $\epsilon$
term in $H_1$ is symmetric, it commutes with $H_0$, and therefore this
kind of perturbation cannot be eliminated by symmetrization. Indeed,
the evolution of the qubit state $\alpha\choose\beta$ is given (if we
ignore the $\eta$ term, for simplicity) by $\alpha(t)=
\alpha(0)\,e^{-i\epsilon t}$ and $\beta(t)=\beta(0)\,e^{i\epsilon t}$.
If there were $R$ qubits, instead of just two, the symmetric part of
the perturbation (which cannot be eliminated by symmetrization) would
have as its coefficient the arithmetic average of the individual
perturbations. If the latter are random and independent, that average
is expected to be smaller than the r.m.s.\ perturbation by a factor
$\sqrt{R}$, and therefore the error probability is reduced by a factor
$R$. No further improvement can be expected.

On the other hand, the error due to the antisymmetric part of $H_1$ can
be considerably reduced. Written with the Bell basis [11], the
initial state of the pair is

\beq {\alpha\choose\beta}\otimes{\alpha\choose\beta}=
 \frac{\alpha^2+\beta^2}{\sqrt{2}}\,\Ph^+ 
 +\frac{\alpha^2-\beta^2}{\sqrt{2}}\,\Ph^- +
 \sqrt{2}\alpha\beta\,\Ps^+,\label{state}\eeq
where $\Ph^+$, $\Ph^-$, and $\Ps^+$ are the triplet states corresponding
to $J_x=0$, $J_y=0$, and $J_z=0$, respectively.
The antisymmetric part of the perturbation has matrix elements given by

\beq (\sigma_{Az}-\sigma_{Bz})\,\Ph^\pm=0, \eeq
and

\beq (\sigma_{Az}-\sigma_{Bz})\,\Ps^\pm=\Ps^\mp, \eeq
where $\Ps^-$ is the singlet state.  The nontrivial part of the
Hamiltonian thus involves only the $\Ps^\pm$ subspace. We can write
(ignoring for simplicity the $\epsilon$ contribution, which is
symmetric)

\beq H=H_0+H_1= \left( \begin{array}{cc} 0 & \eta \\
 \eta & \Omeg \end{array}  \right). \eeq
It is easy to find the eigenvalues and eigenvectors of this Hamiltonian.
The initial state~(\ref{state}) can be written as a linear combination
of these two eigenstates, and its time evolution obtained
explicitly:  the $\Ph^\pm$ terms have constant amplitudes, and, for
$\eta\ll\Omeg$, the $\Ps^+$ term in~(\ref{state}) evolves as

\beq \Ps^+\to e^{i\eta^2t/\Omeg}\,\Ps^+
 +(\eta/\Omeg)\,(e^{-i\Omeg t}-1)\,\Ps^-, \eeq
where terms of order $(\eta/\Omeg)^2$ have been neglected. If we could
make the potential energy $\Omeg$ arbitrarily large (as we do in an
ideal ``quantum measurement'' context, where the interaction with the
measuring apparatus is assumed arbitrarily strong), then the $\Ps^+$
term in the state vector would be perfectly stabilized, and the $\Ps^-$
term (which is anti\-symmetric) would never appear. For large but
finite $\Omeg$, the amplitude of the $\Ps^-$ term, initially zero,
always remains small. On the other hand, the $\Ps^+$ term undergoes a
slow secular drift, which definitely is an error, but is nevertheless
compatible with the symmetry constraint. The same kind of drift also
occurs for repeated discrete symmetrization [9], because the symmetric
state obtained at each step may contain a small residual error, and
these errors gradually accumulate.

These considerations can now be generalized from 2 to $R$ computers,
each one having many qubits. It may seem that a global potential is
required, involving all of them at once, a proposal that would be a
technological nightmare. Fortunately, this is not necessary: it is
enough to take $R(R-1)/2$ identical potentials, one for each pair of
computers. If any two computers are in a symmetric state, then all $R$
computers are in a symmetric state, by definition. However, the
comparison of two computers cannot be done bitwise: the states of the
complete computers have to be compared. How to actually do that by means
of a coherent sequence of two-qubit interactions requires a complicated
protocol~[10], beyond the scope of this review.\\[7mm]

\noindent{\bf 3. Encoding and decoding}\bigskip

Let us now turn our attention to the case of large errors, occurring in
a few, isolated qubits. The latter are materialized by single quanta,
such as trapped ions~[12]. Their coupling to a dissipative environment
(which was the standard stabilizing mechanism for classical bits) is to
be avoided as much as possible, because it readily leads to
decoherence, namely to the loss of phase relationships. Yet,
disturbances due to the environment cannot be completely eliminated:
e.g., even if there are no residual gas molecules in the vacuum of an
ion trap, there still are the vacuum fluctuations of the quantized
electromagnetic field, which induce spontaneous transitions between the
energy levels of the ions.  Therefore, error control is an essential
part of any quantum communication or computing system.

This goal is much more difficult to achieve than classical error
correction, because qubits cannot be read, or copied, or duplicated,
without altering their quantum state in an unpredictable way [13]. The
feasibility of quantum error correction, which for some time had been in
doubt, was first demonstrated by Shor [7]. As in the classical case,
redundancy is an essential element, but this cannot be a simple
repetitive redundancy, where each bit has several identical replicas
and a majority vote is taken to establish the truth.  This is because
qubits, contrary to ordinary classical bits, can be {\it entangled\/},
and usually they are. As a trivial example, in the singlet state of two
spin-$1\over2$ particles, each particle, taken separately, is in a
completely random state. Therefore, comparing the states of
spin-$1\over2$ particles that belong to different (redundant) singlets
would give no information whatsoever.

All quantum error correction methods [7, 8, 14--17] use several
physical qubits for representing a smaller number of logical qubits
(usually a single one). These physical qubits are prepared in a
carefully chosen, highly entangled state. None of these qubits, taken
alone, carries any information. However, a large enough subset of them
may contain a sufficient amount of information, encoded in relative
phases, for determining and exactly restoring the state of the logical
qubits, including their entanglement with the other logical qubits in
the quantum computer.

I shall now review the quantum mechanical principles that make error
correction possible. (I shall not discuss how to actually design new
codewords; the most efficient techniques involve a combination of
classical coding theory and of the theory of finite groups~[17].) Note
that, since quantum codewords span only a restricted subspace of the
complete physical Hilbert space, the unitary operations that generate
the quantum dynamical evolution (that is, the computational process)
are subject to considerable arbitrariness. The latter is similar to the
gauge freedom in quantum field theory. Quantum codewords can thus serve
as a simple toy model for investigating the quantization of constrained
dynamical systems, such as field theories with gauge groups [18].

In the following, I shall usually consider codewords that represent a
single logical qubit. It is also possible, and it is more efficient, to
encode several qubits into larger codewords. Ideally, it would be best
to encode the entire computer into one super-codeword (but then, how to
program the evolution of that super-codeword would be a very difficult
problem, that might be solvable only by another quantum computer!).
However, no new physical principles are involved in the simultaneous
encoding of several qubits, and the simple case of a single qubit is
sufficient for illustrating these principles. 

The quantum state of a single logical qubit will be denoted as

\beq \psi=\alpha\,\0+\beta\,\1, \label{qubit}\eeq
where the coefficients $\alpha$ and $\beta$ are complex numbers. The
symbols \0\ and \1\ represent any two orthogonal quantum states, such as
``up'' and ``down''  for a spin, or the ground state and an excited
state of a trapped ion.  In a quantum computer, there are many logical
qubits, typically in a collective, highly entangled state, and any
particular qubit has no definite state. I shall still use the same
symbol $\psi$ for representing the state of the entire computer, and
Eq.~(\ref{qubit}) could now be written as

\beq \psi=|\alpha\ket\otimes\0+|\beta\ket\otimes\1, \label{computer}\eeq
where one particular qubit has been singled out for the discussion, and
the symbols $|\alpha\ket$ and $|\beta\ket$ represent the collective
states of all the other qubits, that are correlated with \0\ and \1,
respectively.  However, to simplify the notation and improve
readability, I shall still write the computer state as in
Eq.~(\ref{qubit}). In the following, Dirac's ket notation will in
general {\it not\/} be used for generic state vectors (such as $\psi$,
$\alpha$, $\beta$) and the $\otimes$ sign will sometimes be omitted,
when the meaning is clear. Kets will be used only for denoting basis
vectors such as \0\ and \1, and their direct products.  The latter will
be labelled by binary numbers, such as

\beq |9\ket\equiv|01001\ket\equiv\0\otimes\1\otimes\0\otimes\0\otimes\1. 
\eeq

In order to encode the qubit $\psi$ in Eq.~(\ref{qubit}), we intoduce
an auxiliary system, called {\it ancilla\/} (this is the Latin word for
housemaid). The ancilla is made of $n$ qubits, initially in a state
$|000\ldots\,\ket$. We shall use $2^n$ mutually orthogonal
vectors $|a\ket$, with $a=0$,~1, \ldots\ (written in binary notation),
as a basis for the quantum states of the ancilla. The labels $a$ are
called {\it syndromes\/}, because, as we shall see, the presence of an
ancilla with $a\neq0$ may serve to identify an error in the
encoded system that represents $\psi$.

Encoding is a unitary transformation, $E$, performed on a physical qubit
and its ancilla together:

\beq \zz\otimes\ao\to E\,\Bigl(\zz\otimes\ao\Bigr)\equiv|Z_0\ket,\eeq
where $z$ and $Z$ are either 0 or 1 (the index 0 in $|Z_0\ket$ means
that there is no error at this stage). The unitary transformation $E$
is executed by a quantum circuit (an array of quantum gates). However,
from the theorist's point of view, it is also convenient to consider
$\zz\otimes\ao$ and $|Z_0\ket$ as two different representations of the
same qubit \zz: its logical representation, and its physical
representation. The first one is convenient for discussing matters of
principle, such as quantum algorithms, while the physical
representation shows how qubits are actually materialized by distinct
physical systems (which may be subject to {\it independent\/} errors).
These two different representations are analogous to the use of normal
modes vs.\ local coordinates for describing the small oscillations of a
mechanical system [19]. One description is mathematically simple, the
other one refers to directly accessible quantities.\\[7mm]

\noindent{\bf 4. Error correction}\bigskip

\noindent Since there are $2^n$ syndromes (including the null syndrome
for no error), it is possible to identify and correct up to $2^n-1$
different errors that affect the physical qubits, with the help of a
suitable decoding method, as explained below. Let $|Z_a\ket$, with
$a=0$, \ldots\,, $2^n-1$, be a complete set of orthonormal vectors
describing the physical qubits of which the codewords are made:
$|0_0\ket$ and $|1_0\ket$ are the two error free states that represent
\0\ and \1, and all the other $|0_a\ket$ and $|1_a\ket$ are the results
of errors (affecting one physical qubit in the codeword, or several
ones, this does not matter at this stage).  These $|Z_a\ket$ are
defined in such a way that $|0_a\ket$ and $|1_a\ket$ result from
definite errors in the same physical qubits of $|0_0\ket$ and
$|1_0\ket$:  for example, the third qubit is flipped,
${\alpha\choose\beta}\to{\beta\choose\alpha}$, and the seventh one has
a phase error, ${\alpha\choose\beta}\to{\alpha\choose -\beta}$. 

We thus have two complete orthonormal bases, $\zz\otimes|a\ket$ and 
$|Z_a\ket$.  These two bases uniquely define a unitary transformation
$E$, such that

\beq E\,(\zz\otimes|a\ket)=|Z_a\ket, \eeq
and

\beq E^\dagger\,|Z_a\ket=\zz\otimes|a\ket, \label{decod} \eeq
where $a$ runs from 0 to $2^n-1$. Here, $E$ is the encoding matrix, and
$E^\dagger$ is the decoding matrix. If the original and corrupted
codewords are chosen in such a way that $E$ is a real orthogonal matrix
(not a complex unitary one), then $E^\dagger$ is the transposed matrix,
and therefore $E$ and $E^\dagger$ are implemented by the {\it same\/}
quantum circuit, executed in two opposite directions. (If $E$ is
complex, the encoding and decoding circuits must also have opposite phase
shifts.)

The $2^n-1$ standard errors $|Z_0\ket\to|Z_a\ket$ are not the only
ones that can be corrected by the $E^\dagger$ decoding. Any error of
type

\beq |Z_0\ket\to U\,|Z_0\ket=\sum_a c_a\,|Z_a\ket, \eeq
is also corrected, since

\beq E^\dagger\,\sum_a c_a\,|Z_a\ket=\zz\otimes\sum_a c_a\,|a\ket, \eeq
is a direct product of \zz\ with the ancilla in some irrelevant
corrupted state. Note that {\it no knowledge of the syndrome is
needed\/} in order to correct the error [6]. Error correction is a
logical operation that can be performed automatically, without having
to execute quantum measurements. We definitely know that the error is
corrected, even if we don't know the nature of that error. I shall
return to this issue in Sect.~5.

It is essential that the result on the right hand side of (\theequation)
be a direct product. Only if the new ancilla state is the same for
$\zz=\0$ and $\zz=\1$, and therefore also for the complete computer
state in Eq.~(\ref{computer}), is it possible to coherently detach
the ancilla from the rest of the computer, and replace it by a fresh
ancilla (or restore it to its original state \ao\ by a dissipative
process involving still another, extraneous, physical system). This
means, in the graphical formalism of quantum circuits, that the
``wires'' corresponding to the old ancilla stop, and new ``wires''
enter into the circuit, with a standard quantum state for the new
ancilla. There is some irony in this introduction of a dissipative
process for stabilizing a quantum computer. The latter was originally
conceived as an analog device with a continuous evolution, and it is
now brought one step closer to a conventional digital computer!

There are many plausible scenarios for the emergence of coherent
superpositions of corrupted states, as in (\theequation). For example,
in an ion trap, a residual gas molecule, whose wave function is spread
over a domain much larger than the inter-ion spacing, can be scattered
by all the ions together, as by a diffraction grating, and then all the
ions are left in a collective recoil state (namely, a coherent
superposition of states where one of the ions recoiled and the other
ones did not).  Furthermore, {\it mixtures\/} of errors of type
(\theequation) are also corrigible. Indeed, if

\beq \rho=\sum_j p_j\,\sum_{ab}c_{ja}\,|Z_a\ket\,\bra Z_b|\,c^*_{jb},
\eeq
with $p_j>0$ and $\sum p_j=1$, then

\beq E^\dagger\rho\,E= \zz\,\bra z|\otimes
  \sum_j p_j\,\sum_{ab}c_{ja}\,|a\ket\,\bra b|\,c^*_{jb}, \eeq
again is a direct product of the logical qubit and the corrupted
ancilla in a mixed state.

In particular, these mixtures include the case where a physical qubit
in the codeword gets entangled with an unknown environment, which is
the typical source of error. Let $\eta$ be the initial, unknown state
of the environment, and let its interaction with a physical qubit cause
the following unitary evolution:

\beq \begin{array}{lll}
 \0\otimes\eta & \to & \0\otimes\mu+\1\otimes\nu,\smallskip\\
 \1\otimes\eta & \to & \0\otimes\sigma+\1\otimes\tau,\end{array}
 \label{error} \eeq
where the new environment states $\mu,\ \nu,\ \sigma$, and $\tau$, are
also unknown, except for unitarity constraints. Now assume that the
physical qubit, which has become entangled with the environment in such
a way, was originally part of a codeword,

\beq |Z_0\ket=|X_{Z0}\ket\otimes\0+|X_{Z1}\ket\otimes\1, \eeq
where the index $Z$ means 0 or 1. (The index 0 may also refer to the
error free state of a codeword. The interpretation of a subscript 0
should be obvious from the context.) The codeword $|Z_0\ket$, together
with its environment, thus evolves as

\beq Z_0\otimes\eta\to Z'=X_{Z0}\2\Bigl(\0\2\mu+\1\2\nu\Bigr)
  +X_{Z1}\2\Bigl(\0\2\sigma+\1\2\tau\Bigr), \eeq
where I have omitted most of the ket signs, for
brevity. This can also be written as\vspace{-2mm}

\beq \begin{array}{lll}
 Z' & = & 
  \Bigl[X_{Z0}\2\0+X_{Z1}\2\1\Bigr]\,{\displaystyle{\mu+\tau\over2}}+
  \Bigl[X_{Z0}\2\0-X_{Z1}\2\1\Bigr]\,{\displaystyle{\mu-\tau\over2}}\;+
  \smallskip \\ & &
  \Bigl[X_{Z0}\2\1+X_{Z1}\2\0\Bigr]\,{\displaystyle{\nu+\sigma\over2}}+
  \Bigl[X_{Z0}\2\1-X_{Z1}\2\0\Bigr]\,{\displaystyle{\nu-\sigma\over2}}\,.
  \end{array}\eeq
On the right hand side, the vectors

\beq \begin{array}{lll}
 Z_0 & := & X_{Z0}\2\0+X_{Z1}\2\1,\smallskip \\
 Z_r & := & X_{Z0}\2\0-X_{Z1}\2\1,\smallskip \\
 Z_s & := & X_{Z0}\2\1+X_{Z1}\2\0,\smallskip \\
 Z_t & := & X_{Z0}\2\1-X_{Z1}\2\0,\end{array} \label{Z}\eeq
correspond, respectively, to a correct codeword, to a phase error
($\1\to-\1$), a bit error ($\0\leftrightarrow\1$), which is the only
classical type of error, and to a combined phase and bit error.  If
these three types of errors can be corrected, we can also correct any
type of entanglement with the environment, as we shall soon see.

For this to be possible, it is sufficient that the eight vectors in
Eq.~(\ref{Z}) be mutually orthogonal. The simplest way of achieving this orthogonality is to
construct the codewords $|0_0\ket$ and $|1_0\ket$ in such a way that
the following scalar products hold:

\beq \bra X_{Zy}\,,\,X_{Z'y'}\ket=
 \mbox{$1\over2$}\,\delta_{ZZ'}\,\delta_{yy'}. \label{XX}\eeq
(There are 10 such scalar products, since each index in this equation
may take the values 0 and 1.) If these conditions are satified, the
decoding of $Z'$ by $E^\dagger$ gives, by virtue of Eq.~(\ref{decod}),

\beq E^\dagger\,Z'=\zz\otimes\left(\ao\2{\mu+\tau\over2}
 +|r\ket\2{\mu-\tau\over2}+|s\ket\2{\nu+\sigma\over2}
 +|t\ket\2{\nu-\sigma\over2}\right),\eeq
where $|r\ket$, $|s\ket$, and $|t\ket$ are various corrupted states of
the ancilla. The expression in parentheses is an entangled state of
the ancilla and the unknown environment. We cannot know it explicitly,
but this is not necessary: it is sufficient to know that it is the same
state for $\zz=\0$ and $\zz=\1$, and any linear combination thereof, as
in Eq.~(\ref{qubit}). We merely have to discard the old ancilla and
bring in a new one. 

This is how codewords fight entanglement with entanglement. How to 
actually construct codewords that satisfy Eq.~(\ref{XX}), when
{\it any\/} one of their physical qubits is singled out, is a difficult
problem, best handled by a combination of classical codeword theory and
finite group theory [17]. I shall not enter into this subject here. I
only mention that in order to correct an arbitrary error in any one of
its qubits, a codeword must have at least five qubits: each one
contributes three distinct vectors, like $Z_r$, $Z_s$, and $Z_t$  in
Eq.~(\ref{Z}), and these, together with the error free vector $Z_0$,
make 16 vectors for each logical qubit value, and therefore $32=2^5$ in
the total. Longer codewords can correct more than one erroneous qubit.
For example, Steane's linear code [15], with 7 qubits, can correct not
only any error in a single physical qubit, but also a phase error,
$\1\to-\1$, in one of them, and a bit error, $\0\leftrightarrow\1$, in
another one (check!  $1+7\times3+7\times6=2^{7-1}$). A well designed
codeword is one where the orthogonal basis $|Z_a\ket$ corresponds to
the most plausible physical sources of errors, e.g., single bit errors,
rather than complicated types of errors involving several qubits in a
coherent way.

The error correction method proposed above, in Eq.~(\ref{decod}), is
conceptually simple, but it has the disadvantage of leaving the logical
qubit \zz\ in a ``bare'' state, vulnerable to new errors that would be
not be detected. It is therefore necessary to re-encode that qubit
immediately, with another ancilla (or with the same ancilla, reset to
\ao\ by interaction with still another system). A more complicated but
safer method is to bring in a second ancilla, in a standard state
$|b=0\ket$, and have it interact with the complete codeword in such a
way that

\beq |Z_a\ket\otimes|b=0\ket\to |Z_0\ket\otimes|b=a\ket. \label{b} \eeq
This again is a unitary transformation, which can be implemented by a
quantum circuit. Note that now the unitary matrix that performs that
error recovery is of order $2^{2n+1}$, instead of $2^{n+1}$.

Naturally, errors can also occur in the encoding and decoding processes
themselves.  More sophisticated methods must therefore be designed,
that allow fault tolerant computation. An adaptive strategy is
indicated, with several alternative paths for error correction. Most
paths fail, because new errors are created; however, these errors can
be detected, and there is a high probability that one of the paths will
eventually lead to the correct result. As a consequence, the error
correction circuits are able to correct old errors faster than they
introduce new ones. There is then a high probability for keeping the
number of errors small enough, so that the correction machinery can
successfully deal with them [20, 21].

Finally, let us note that the symmetrization method, that was discussed
in Sect.~2, can also be applied to individual codewords, if the latter
have an internal symmetry. For example the codewords of ref.~[8] are
invariant under cyclic permutations of their 5 qubits. These codewords
have the property that if any four qubits are correct, it is always
possible to restore the remaining defective qubit.  However, the
codeword error correction procedure definitely requires four qubits to
be correct, and it cannot cope with small drifts of all five qubits, or
even two of them.  Therefore, it is helpful to test once in a while the
cyclic symmetry of the codeword: successful tests will reduce the
amplitude of small errors. Unfortunately, just as in the case of
inter-computer symmetrization, an unsuccessful test leads to an
asymmetric state, and forces us to completely discard the incorrect
codeword. One of the logical qubits is then missing, and the
computation can proceed only if there is enough redundancy among the
logical qubits themselves (not only in their representation by physical
qubits), for example, if they are parts of higher order
codewords.

Instead of continually testing the symmetry of a codeword, it is also
possible to force its physical qubits to respect that symmetry by
introducing a high potential barrier that prevents access to asymmetric
states, as in Eq.~(\ref{Om}):

\beq H_0\to 
  H_0+\Omeg\,\Bigl(1-|0_0\ket\bra0_0|-|1_0\ket\bra1_0|\Bigr).\eeq
In this way, an error that turns a codeword state into a new quantum
state lying in the ortho\-complement of the legal subspace, can be
produced only by investing a large amount of energy, $\Omeg$. All
corrigible errors are of this type, and are therefore prevented.
However, an incorrigible error creates a state that is not orthogonal
to both codewords, and therefore it would not be prevented by the
additional potential in Eq.~(\theequation). Incorrigible errors remain
uncorrected, of course.  At most, their probability of occurrence can
be reduced.\\[7mm]

\noindent{\bf 5. To know or not to know}\bigskip

It was already noted that no knowledge of the syndrome is needed in
order to correct a corrigible error [6]. Error correction is a logical
operation. It is part of the software, and can be performed
automatically, without involving irreversible quantum measurements. We
can be sure that the error is corrected, even if we don't know the
nature of that error.

This situation is reminiscent of the teleportation of an {\it
unknown\/} quantum state~[22]: the classical information sent by the
emitter is not correlated to the quantum state that is teleported, and
the emitter does not know which state she sends. The receiver also does
not know, and cannot know, which state he receives. However, he can be
sure that this state is identical to the one that was in the emitter's
hands, before the teleportation process began. The teleportation
process can also be achieved by an ordinary quantum circuit~[23],
without performing any measurement.  Alternatively, that circuit can be
interrupted by a quantum measurement, and classical information
transferred in a conventional way to another point, where the circuit
restarts. There, the classical information is used for performing a
unitary transformation that brings the teleportation process to
successful completion.

Likewise, there is no fundamental difference between a ``conscious''
error correction and an ``unconscious'' one. Figure~1 describes how a
codeword is measured, in order to determine the syndrome and correct
the error.  On the other hand, Fig.~2(a) shows how the same error
correction can be performed automatically, thanks to Eq.~(\ref{b}). In
that case, the syndrome need not be known, but it still is encoded in 
an ancilla. It may therefore be measured, if we wish so, in order to
restore that ancilla to its standard state, $|b=0\ket$. This is shown
in Fig.~2(b).
From a comparison of these figures, it is clear that automatic
correction is conceptually simpler, and makes use of fewer hardware
resources than those needed for measuring and then correcting an
error.\\[7mm]

\noindent{\bf Acknowledgments}\bigskip

\noindent I am grateful to David DiVincenzo, Peter Shor, and Andrew
Steane for clarifying remarks. This research was supported in part by
the National Science Foundation under Grant No.\ PHY94-07194.\medskip

\bigskip\noindent{\bf References}\frenchspacing

\begin{enumerate}
\setlength{\leftmargin}{-.25in}
\item D. Welsh, {\it Codes and Cryptography\/}, Oxford University Press
(1989), Chapt. 4.
\item A. Peres, \ {\it Quantum Theory: Concepts and Methods\/}, \ Kluwer,
\ Dordrecht (1993), Chapt.~5.
\item P. A. Benioff, J. Stat. Phys. 22 (1980) 563.
\item R. P. Feynman, Found. Phys. 16 (1986) 507.
\item D. Deutsch, Proc. Roy. Soc. (London) A 400 (1985) 97.
\item A. Peres, Phys. Rev. A 32 (1985) 3266.
\item P. W. Shor, Phys. Rev. A 52 (1995) 2493.
\item C. H. Bennett, D. P. DiVincenzo, J. A. Smolin, and W. K. Wootters,
Phys. Rev. A 54 (1996) 3824.
\item A. Berthiaume, D. Deutsch, and R. Jozsa, ``The stabilisation of
quantum computation'', in {\it Proceedings of
 the Workshop on Physics and Computation, PhysComp'94\/}, IEEE Computer
 Society Press (1994), 60--62.
\item A. Barenco, A. Berthiaume, D. Deutsch, A. Ekert, R. Jozsa, and
C. Macchiavello, ``Stabilisation of quantum computations by symmetrisation,''
 e-print quant-ph 9604028.
\item S. L. Braunstein, A. Mann, and M. Revzen, Phys. Rev. Lett. 68
(1992) 3259.
\item J. I. Cirac and P. Zoller, Phys. Rev. Lett. 74 (1995) 4091.
\item W. K. Wootters and W. H. Zurek, Nature 299 (1982) 802.
\item R. Laflamme, C. Miquel, J. P. Paz, and W. H. Zurek, Phys. Rev.
Lett. 77 (1996) 198.
\item A. M. Steane, Phys. Rev. Lett. 77 (1996) 793; Proc. Roy. Soc.
(London) A 452 (1996) 2551.
\item E. Knill and R. Laflamme, ``A theory of quantum error-correcting
codes'' (Los Alamos report LA-UR-96-1300).
\item A.~R. Calderbank, E.~M. Rains, P.~W. Shor, and
N.~J.~A. Sloane, ``Quantum Error Correction and Orthogonal Geometry,''
e-print quant-ph 9605005; ``Quantum Error Correction via Codes over
GF(4),'' e-print quant-ph 9608006, submitted to IEEE Transactions on
Information Theory.
\item A. Peres, ``Unitary dynamics for quantum codewords'' in {\it Quantum
Communication, Computing and Measurement\/}, ed. by O.~Hirota {\it et
al.\/} (Plenum, 1997).
\item H. Goldstein, {\it Classical Mechanics\/}, Addison-Wesley, Reading
(1980), Chapt. 6.
\item P. W. Shor, ``Fault tolerant quantum computation'' in {\it Proc.
37th Symposium on Foundations of Computer Science\/} (1996) in press.
\item D. Aharonov and M. Ben-Or, ``Fault tolerant quantum computation
with constant error'', e-print quant-ph 9611025.
\item C. H. Bennett, G. Brassard, C. Cr\'epeau, R. Jozsa, A. Peres, and
W. K. Wootters, Phys. Rev. Lett. 70 (1993) 1895.
\item G. Brassard, ``Teleportation as a quantum computation'' in {\it
Fourth Workshop on Physics and Computation\/} ed. by T. Toffoli, M.
Biafore and J. Le\~ao (New England Complex Systems Institute, 1996),
48--50.

\end{enumerate}\vfill

\noindent {\bf Captions of figures}\bigskip

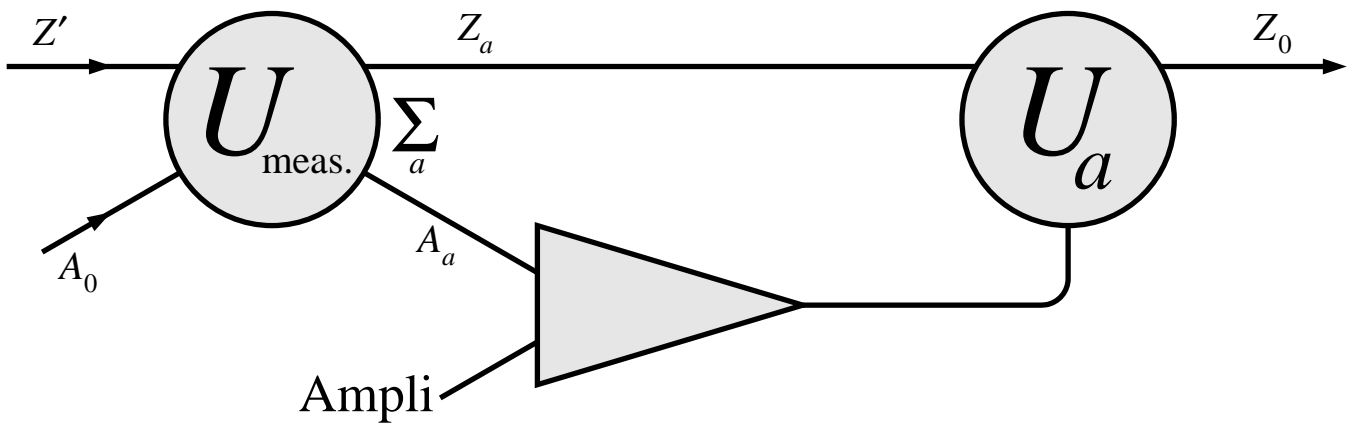
\noindent Fig.~1. \ The corrupted codeword $|Z'\ket$ interacts with a
measuring apparatus, initially in state $|A_0\ket$. The codeword and the
apparatus get entangled (their states $|Z_a\ket$ and $|A_a\ket$
are correlated). Amplification brings the apparatus into the classical
domain, and gives a definite value to $a$. The apparatus then selects
the unitary transformation $U_a$ that restores the correct codeword
state, $|Z_0\ket$.\bigskip

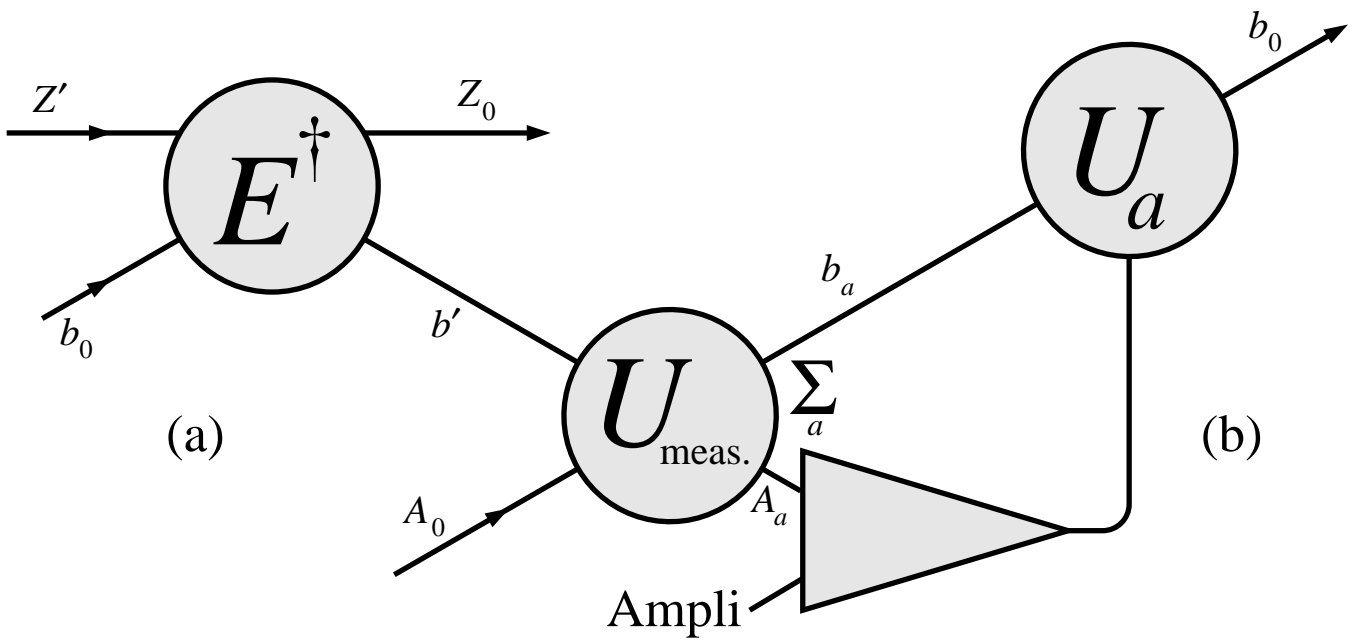
\noindent Fig.~2. \ (a) The corrupted codeword $|Z'\ket$ interact with a
fresh ancilla $|b_0\ket$, and returns to the legal state $|Z_0\ket$,
while the used ancilla in state $|b'\ket$ may be discarded. \ (b)
Instead of discarding the ancilla, it is possible to measure it by means
of an apparatus, initially in state $|A_0\ket$. The ancilla and the
apparatus then get entangled, and the restoration process continues as in
Fig.~1.

\end{document}